\documentclass{ws-ijmpa}

\newcommand{\de}{\mathrm{d}}

\newcommand{\Sym}{\mathrm{Sym}}
\newcommand{\End}{\mathrm{End}}

\begin{document}

\markboth{Antonio Ricco}
{Brane superpotential and local Calabi--Yau manifolds}

%
\catchline{}{}{}{}{}
%

\title{BRANE SUPERPOTENTIAL \\
AND LOCAL CALABI--YAU MANIFOLDS}

\author{ANTONIO RICCO}
\address{Via dei Volsci, 10,  
 	 00185 Roma,
         Italy\\
         anricco@gmail.com}

\maketitle


\begin{abstract}
We briefly report on some recent progress in the computation of 
B-brane superpotentials for Type II strings compactified on Calabi--Yau manifolds,
obtained by using a parametrization of tubular neighborhoods of complex submanifolds, 
also known as local spaces. 
In particular, we propose a closed expression for the superpotential of a brane 
on a genus-$g$ curve in a Calabi--Yau threefold in the case in which there exists 
a holomorphic projection from the local space around the curve to the curve itself.

\end{abstract}

\ccode{PACS numbers: 11.25.Hf, 123.1K}

\section{Introduction}

Interesting $\mathcal{N}=1$ gauge theories can be obtained as low-energy 
limits of Type II string theories compactified on Calabi--Yau manifolds 
with internal boundary conditions on holomorphic submanifolds. 
The tree level superpotential of such theories 
corresponds to the disk amplitudes of the topological B-model, and, in principle,  
can be computed in various ways, e.g. by using worldsheet techniques, by reducing the
holomorphic Chern--Simons theory to the brane world-volume, by using an appropriate
locally-free resolution, by considering the Landau-Ginzburg phase of the
model, or by various combinations of those methods, see Refs. \refcite{Baumgartl}--\refcite{todorov}
and the references therein.
The method that we present here exploits the relation 
between the superpotential, in the large radius limit and when 
the ambient Calabi--Yau manifold corresponds to the local space, and the 
deformation theory of the submanifold,
along the lines of Ref.~\refcite{katz}.
It has the advantage of being reasonably pratical for explicit computations and steems from
the remark that the deformation theory of a submanifold $M$ with given normal bundle $V$
is intrinsic to the couple $(M, V)$.

\section{Tubular neighborhoods and deformations}
We start by reviewing some properties of
compact complex submanifolds.
Let $M$ be a 
compact complex submanifold of a 
complex manifold $X$, with normal bundle $V$.
Then, for any $|\epsilon| > 0$, there exists a tubular neighborhood%
\footnote{The name \emph{local spaces} for such tubular neighborhoods can be also understood as a generalizazion of the 
``local'' description of a Riemannian structure around a point.}
$X_M$ of $M$ in $X$ and an open covering  $\mathcal{U}=\{U_\alpha\}$ of $M$ such that $X_M$ can be covered
by an atlas $W_\alpha:= U_\alpha \times D^r$ ($D:=\{w \in \mathbb{C}: |w| < 1\}$), with transition functions
\begin{eqnarray}
   \left(\begin{array}{c}
          z_{\alpha} \\
          w_{\alpha}
         \end{array} \right) =    
   \left(\begin{array}{c}
          f_{\alpha \beta} (z_\beta) \\
          V_{\alpha\beta} (z_\beta)  w_{\beta}
         \end{array} \right) +
   \epsilon M_{\alpha\beta} (z_\beta, w_\beta)
   \left(\begin{array}{c}
           \zeta_{\alpha\beta} \\
           \omega_{\alpha \beta}
         \end{array} \right) 
\end{eqnarray}
where $\{f_{\alpha \beta}\}$ are the transition functions for $M$, $\{ V_{\alpha \beta}\}$ and 
 $\{ M_{\alpha\beta}  \}$ those for the normal bundle $V$ and its tangent space $T_V$ respectively.
If we ignore convergence issues and treat the deformation terms 
$\{(\zeta_{\alpha\beta}, \omega_{\alpha\beta})\}$ 
as polynomials%
\footnote{Thus, the following considerations should be made for the appropriate completions of the spaces considered.}
in the $\{w_\alpha\}$ then\cite{Ricco2} the deformation terms
are cocycles representing elements in the \v{C}ech cohomology group ${H}^1(V, T_V)$, 
up to terms of order $\epsilon^2$.
In particular, the space 
$H^1(M, V \otimes \Sym V^*)$ parametrizes the tubular neighborhoods that admit
a projection to the base $M$.
The tubular neighborhoods of complex submanifolds of Calabi--Yau 
manifolds are those with $\det V \simeq K_M$ and are parametrized by the kernel of a map
induced in cohomology by the \emph{divergence map} acting on the sections of the 
tangent bundle.

Now, given a compact submanifold $M$ of a complex manifold $X$, there exists\cite{namba}
a map, 
$K : U \to H^1 (M, V)$, where $U \subset H^0(M,V)$ is a neighborhood of $0$,
with $K(0) = 0$, $\de K (0) = 0$ and 
such that $K^{-1} (0)$ parametrizes the local deformations of $M$ in $X$. 
The map $K$ only depends on the local proprierties of the embedding, i.e. only on a 
tubular neighborhood of $M$ in $X$.
Up to convergence issues,%
\footnote{Notice, however, that the convergence properties of the map $K$ depend on the 
 the convergence properties of the deformation terms.}
the map $K$ can be seen as an element of
\begin{eqnarray}\label{eq:elementK}
  K &\in & H^1(M, V) \otimes \Sym H^0(M, V)^* \nonumber \\
 & \simeq& ( H^{m-1} (M, K_M \otimes V^*) \otimes \Sym H^0 (M, V) )^* 
\end{eqnarray}
where in the last line we used Serre duality, and $m = \dim_\mathbb{C} M$.

Summarizing, to a given embedding $M \subset X$, we can associate 
an element $\alpha \in H^1(V, T_V)$ and a map $K$ as in (\ref{eq:elementK}), 
and we would like to determine the relation between the two. 
In the case in which there exists a projection to the base, 
it is very natural to
guess that the map associating $K[\alpha]$ to $\alpha$ is given by the dual of the
multiplication of the sections morphism
\begin{eqnarray}\label{proposal1}
H^{m-1} (M, K_M \otimes V^*) \otimes \Sym H^0 (M, V) \to H^{m-1} (M, K_M \otimes V^{*} \otimes \Sym V ) . 
\end{eqnarray}

\section{The superpotential}

Let now $C$ be a curve in a Calabi--Yau threefold $X$ with normal bundle $V$
and let us consider a \emph{simple} brane wrapped around $C$,
i.e. a vector bundle on $E \to C$ such that $\End E \simeq \mathcal{O}_C$.
From a geometric viewpoint, the superpotential is a map 
\begin{eqnarray}
  W : H^0 (C, V) \to \mathbb{C}
\end{eqnarray}
whose differential $\de W$ gives the map $K$. 
As we did for the map $K$, we can consider $W$ as an element of $\Sym H^0 (C, V)^*$. 
By using the Calabi--Yau condition, one finds that a tubular neighborhood of $C$ in $X$,
when a projection to the base does exist, is 
parametrized by an element in $H^0(C, \Sym V)^*$.
Then, the expression (\ref{proposal1}) implies that the superpotential 
is given by the map
\begin{eqnarray}\label{superpotential}
\mu^*: H^0(C, \Sym V)^* \to \Sym H^0 (C, V)^*
\end{eqnarray}
that is dual to the multiplication of the sections map.
The formula (\ref{superpotential}) seems to be the simplest 
closed expression for the superpotential in this generality,
and reduces to the previously known expressions\cite{Ferrari:2003vp}
for the Laufer rational curve.
It is suitable for explicit computations when the multiplication morphism
is known, for example
in the case of a rational curve $C\simeq \mathbb{P}^1$ or
of an elliptic curve $C \simeq E_\tau$.
In the case in which the brane is given by a general bundle $E$,
the superpotential gives the coupled deformations of the curve $C$
and of the vector bundle $E$, and  should be obtained by generalizing (\ref{superpotential}), 
although the correct guess is not completely trivial.

\section*{Acknowledgments}

Part of the present work was done with support from Fondazione Angelo della Riccia, Firenze.
The author would also like to thank IMPA, Instituto Nacional de Matem\'atica Pura e Aplicada, 
in Rio de Janeiro, for hospitality.


\end{document}